\documentstyle{aipproc}
\begin{document}
\title{Scalar Mesons as ``Simple'' $Q\bar{Q}$ States}
\author{Eef van Beveren$^*$ and George Rupp$^{\dagger}$}
\address{$^*$Departamento de F\'{\i}sica, Universidade de Coimbra, 3000
Coimbra, Portugal\\ (eef@teor.fis.uc.pt) \\
$^{\dagger}$Centro de F\'{\i}sica das Interac\c{c}\~{o}es Fundamentais,
Instituto Superior T\'{e}cnico, Edif\'{\i}cio Ci\^{e}ncia, 1049-001 Lisboa,
Portugal (george@ajax.ist.utl.pt)}

%\lefthead{LEFT head}
%\righthead{RIGHT head}
\maketitle

\begin{abstract}
The Nijmegen Unitarized Meson Model and its application to scalar mesons is
briefly revisited. It is shown that all scalar states up to 1.5 GeV can
be described as $^3P_0$ $q\bar{q}$ states coupled to the OZI-allowed open and
closed two-meson channels consisting of pseudoscalar and vector mesons. Crucial
are the manifestation of a resonance-doubling phenomenon, typical for strong
$S$-wave decay, and the employment of truly flavor-symmetric coupling
constants. Also $S$-wave meson-meson scattering is thus reasonably well
described, without any parameter fit.
\end{abstract}

\section*{Introduction}
The scalar mesons have been posing a serious problem to hadron spectroscopists
over the past three decades, since it seems to be impossible to group these
particles into standard nonets typical for mesonic $q\bar{q}$ systems. Among
the various apparent inconsistencies with standard mesonic states, we should
mention the enigmatic light and broad $\sigma$ meson alias $f_0$(400-1200),
the light and narrow $f_0$(980) (old $S^*$) and $a_0$(980) (old $\delta$), and
the excess of experimental candidates to constitute one ground-state scalar
nonet.  Therefore, a variety of alternative descriptions and mechanisms have
been proposed, such as multiquark ($q^2\bar{q}^2$) configurations, glueballs,
$K\bar{K}$ molecules, and instanton contributions.

In this talk, we shall demonstrate that no such exotic approaches are needed
to obtain a satisfactory description of the scalar-meson sector, provided
one works in a unitarized framework such as the Nijmegen unitarized meson model
(NUMM) \cite{B83}. In particular, the doubling of resonance poles for the
ground states as predicted by the NUMM \cite{B86,B99b} , which is typical for
$S$-wave scattering channels strongly coupled to confined channels \cite{B84},
allows for \em two complete \em \/ scalar nonets, thus accommodating all
experimentally observed states up to 1.5 GeV. Such a resonance doubling was
recently also observed in Refs.~\cite{T95,TR96}, employing a revised version of
the Helsinki unitarized quark model (HUQM). However, in the latter work this
doubling only occurs for some states, which precludes describing, for instance,
an as yet to be confirmed light $K_0^*$ (old $\kappa$) and the established
$f_0$(1500) resonance within the very same framework, contrary to the NUMM.
We ascribe this failure to the use of coupling constants for the three-meson
vertices that are not flavor independent \cite{B99b,B99c}, thus leading to a
breaking of the usual nonet pattern for mesons.

In the following, we shall very briefly review the essence of the NUMM, present
the results for the scalar mesons, and make some concluding remarks in a
perspective of future work.
\section*{Nijmegen unitarized meson model}
The basic unitarization philosophy underlying the NUMM stems from the
observation that most mesons are resonances, some of which so broad that their
very existence seems doubtful, as for example the $f_0$(400-1200). So it makes
no sense to treat such states as stable $q\bar{q}$ systems, even if \em a
posteriori \em \/ hadronic decays are dealt with in perturbation theory. The
problem is that such an approach ignores the possible real mass shifts due to
strong decay, which, at least in principle, could be of the same order of
magnitude as the resonance widths. To make things worse, there is no reason to
presume beforehand that the effects of closed thresholds, corresponding to \em
virtual \/ \em two-meson decays or, in diagrammatic language, to \em mesonic
loops, \em are negligible. 

In order to meet these objections, in the NUMM the valence $q\bar{q}$ system
describing a stable or ``bare'' meson and the various OZI-allowed two-meson
decay channels are treated on an equal footing. To achieve this, use is made of
a coupled-channel Schr\"{o}dinger-type  formalism, in which a physical meson is
represented by a long state vector, for example in the case of the $f_0$ meson
given by
\begin{equation}
|f_0> \;\; = \;\; \left( \begin{array}{cc} n\bar{n} & (l=1) \\ s\bar{s} & (l=1)
\\ \pi\pi & (l=0) \\ \eta_n\eta_n & (l=0) \\\eta_s\eta_s & (l=0) \\  KK & (l=0)
\\ \rho\rho & (l=0) \\ \rho\rho & (l=2) \\ \omega\omega & (l=0) \\ \omega\omega
& (l=2) \\ \phi\phi & (l=0) \\ \phi\phi & (l=2) \\ K^*K^* & (l=0) \\ K^*K^* &
(l=2) \end{array} \right) \;\;\;\;,\;\;\;\; \left\{ \begin{array}{l}
 V_{q\bar{q}} = \frac{1}{2}\mu_q\omega^2r^2 \\[1mm]
 V_{M_1M_2} = 0 \\[1mm]
 V_{qM} = \tilde{g} \frac{r}{r_0} e^{-\frac{1}{2}(\frac{r}{r_0})^2}
\end{array} \right..
\end{equation}
Since the $f_0$ is a scalar isosinglet, we take two $P$-wave $q\bar{q}$
channels that can mix, coupled to a series of $S$- and $D$-wave two-meson
channels consisting of pseudoscalar and vector mesons. The mixing takes place
via the meson-meson channels to which both $n\bar{n}$ and $s\bar{s}$ couple,
i.e., the channels with kaons. Note that $n$ is shorthand for $u$ or $d$
quark, so $\eta_n$ and $\eta_s$ stand for the ideally mixed, non-strange and
strange isosinglet pseudoscalar, respectively. In Eq.~1, we have also
given the used potentials, which amount to a harmonic oscillator with constant
frequency in the $q\bar{q}$ channels, and a peaked function vanishing at the
origin for the transitions between $q\bar{q}$ and meson-meson channels, which
should mimic the $^3P_0$ mechanism (see Ref.~\cite{B83} for reasons and
details). No direct interaction between the two $q\bar{q}$ channels is assumed,
so the mixing takes place via the mesonic channels that couple to both. Also,
no meson-meson or ``final-state'' interactions are included for simplicity.
These assumptions are not strictly necessary, but facilitate the numerical
tractability of the equations \cite{B83} and, moreover, allow a cleaner view
on what are the pure unitarization effects. A possible relaxation of these
restrictions will be discussed in the concluding remarks. As to the generic
coupling constant $\tilde{g}$ in Eq.~(1), it should be noted that it includes
phenomenological factors for one-gluon exchange and closed-threshold
suppression \cite{B83}, besides flavor-symmetric coupling constants for the
various $^3P_0$ three-meson vertices \cite{B99c}.

The NUMM has been applied to heavy quarkonia \cite{B80}, pseudoscalar and
vector mesons \cite{B83}, and scalar mesons \cite{B86}, with generally good
results for the mesonic spectra and meson-meson phase shifts. Especially the
predictions in the scalar sector are of a remarkable quality, considering that
all model parameters have been previously fixed in a fit  to the light and
heavy pseudoscalar and vector mesons, without any additional adjustment to the
scalars. However, before examining in detail the scalar sector, let us first
try to get a feeling for the possible effects of unitarization in a very simple
toy model (see also Ref.~\cite{B91}).

\begin{figure}[h]
\normalsize
\begin{center}
\begin{picture}(283.46,151.73)(-50.00,-30.00)
%
%
%
%
\put(11.16,-5.52){\makebox(0,0)[tc]{3}}
%
\put(66.95,-5.52){\makebox(0,0)[tc]{4}}
%
\put(122.75,-5.52){\makebox(0,0)[tc]{5}}
%
\put(178.54,-5.52){\makebox(0,0)[tc]{6}}
%
%
%
%
%
\put(-5.52,21.57){\makebox(0,0)[rc]{5}}
%
\put(-5.52,43.14){\makebox(0,0)[rc]{10}}
%
\put(-5.52,64.70){\makebox(0,0)[rc]{15}}
%
\put(-5.52,86.27){\makebox(0,0)[rc]{20}}
%
%
%
%
%
%
%
%
%
%
%
%
%
%
%
%
%
\put(234.34,-5.52){\makebox(0,0)[tr]{E}}
\put(113.16,-20.07){\makebox(0,0)[tc]{2 particle CM-energy}}
\put(-5.52,108.14){\makebox(0,0)[tr]{$\sigma$}}
\put(5.52,104.12){\makebox(0,0)[tl]{cross section}}
\put(80.90,0.00){\makebox(0,0){$\times$}}
\put(136.70,0.00){\makebox(0,0){$\times$}}
\put(192.49,0.00){\makebox(0,0){$\times$}}
%
\end{picture}
\end{center}
\normalsize
\caption[]{Elastic scattering cross section for small transition
strength (arbitrary units). \\
Figure reprinted from: E.\ van Beveren, \em Scalar mesons as $q\bar{q}$ systems
with meson-meson admixtures, \em Nucl.\ Phys.\ B (Proc.\ Suppl.) {\bf21}, Page
no.\ 44, Copyright (1991), with permission from Elsevier Science.}
\label{crsmll}
\end{figure}
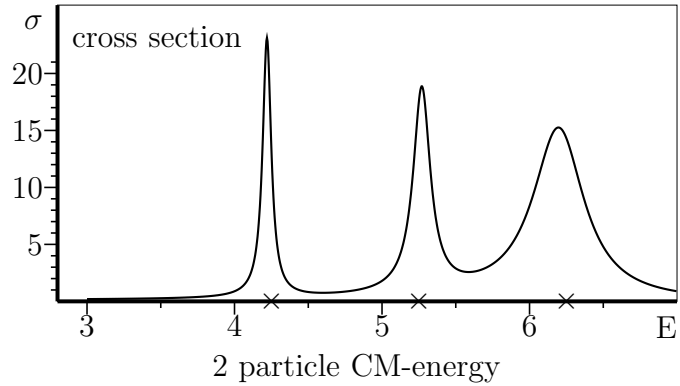

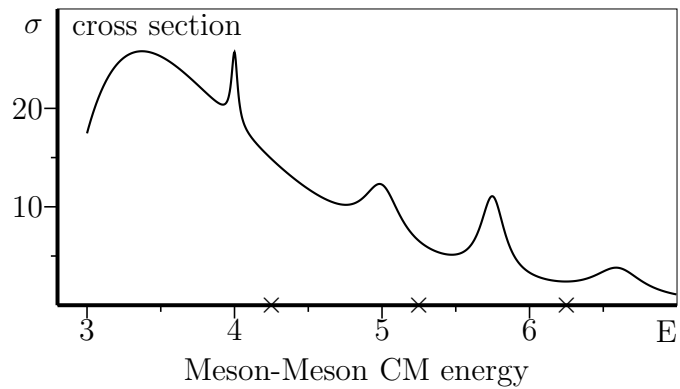
\begin{figure}[h]
\normalsize
\begin{center}
\begin{picture}(283.46,151.73)(-50.00,-30.00)
%
%
%
%
\put(11.16,-5.52){\makebox(0,0)[tc]{3}}
%
\put(66.95,-5.52){\makebox(0,0)[tc]{4}}
%
\put(122.75,-5.52){\makebox(0,0)[tc]{5}}
%
\put(178.54,-5.52){\makebox(0,0)[tc]{6}}
%
%
%
%
%
\put(-5.52,37.26){\makebox(0,0)[rc]{10}}
%
\put(-5.52,74.52){\makebox(0,0)[rc]{20}}
%
%
\put(234.34,-5.52){\makebox(0,0)[tr]{E}}
\put(113.16,-20.07){\makebox(0,0)[tc]{Meson-Meson CM energy}}
\put(-5.52,108.14){\makebox(0,0)[tr]{$\sigma$}}
\put(5.52,110.12){\makebox(0,0)[tl]{cross section}}
\put(80.90,0.00){\makebox(0,0){$\times$}}
\put(136.70,0.00){\makebox(0,0){$\times$}}
\put(192.49,0.00){\makebox(0,0){$\times$}}
%
\end{picture}
\end{center}
\normalsize
\caption[]{Elastic cross section for large transition strength
(arbitrary units). \\
Figure reprinted from: E.\ van Beveren, \em Scalar mesons as $q\bar{q}$ systems
with meson-meson admixtures, \em Nucl.\ Phys.\ B (Proc.\ Suppl.) {\bf21}, Page
no.\ 44, Copyright (1991), with permission from Elsevier Science.}
\label{crlrg}
\end{figure}

To study qualitatively the influence of mesonic decay on the bare spectrum of a
confining $q\bar{q}$ potential, let us consider the two-channel Schr\"{o}dinger
equation
\begin{equation}
\left( \begin{array}{cc}
H_{q\bar{q}}   &  \lambda V_{qM} \\
\lambda V_{qM} & H_{MM}
\end{array} \right) 
\left( \begin{array}{l}
\Psi_{q\bar{q}} \\ \Psi_{MM}
\end{array} \right) = E
\left( \begin{array}{l}
\Psi_{q\bar{q}} \\ \Psi_{MM}
\end{array} \right) \; .
\end{equation}
Here, $H_{q\bar{q}}$ contains a confining potential, that is, a harmonic
oscillator, $H_{MM}$ is taken to be a free $S$-wave Hamiltonian, and $\lambda
V_{qM}$ simulates the transitions between the $q\bar{q}$ and meson-meson
channel through $^3P_0$ quark-pair creation. In the case that the transition
strength $\lambda=0$, Eq.~(2) just yields two disconnected spectra, a discrete
one for the bare $q\bar{q}$ state, and a continuous one for the free two-meson
system. Once $\lambda\neq0$, one unique spectrum emanates, which amounts to a
number of resonances resulting from the possibility for the confined $q\bar{q}$
system to decay into two mesons. In Fig.~\ref{crsmll} \cite{B91}, we plot the
total meson-meson cross section for the case of small $\lambda$. We clearly see
that there is an evident correspondence between the found peaks and the
discrete $q\bar{q}$ energy levels indicated with crosses, the central resonance
positions almost exactly coinciding with the bound-state energies. This is a
situation one typically would find in atomic physics. However, in hadronic
physics the state of affairs is usually very different, where the now large
coupling $\lambda$ reflects the possibility of strong decay. Such a situation
is depicted in Fig.~\ref{crlrg} \cite{B91}, where the resonance peaks and bumps
are not only at energies quite off the values in the bare spectrum, but even
different in number. It is obvious that, if anything similar were to happen in
real hadron spectroscopy, the consequences would be dramatic, since then hardly
any inference could be drawn from the physical spectrum concerning the
underlying confining $q\bar{q}$ potential. In the following application of the
NUMM to scalar mesons, we shall verify that this indeed occurs.

\clearpage

\section*{Scalar Mesons}
As mentioned before, the application of the full NUMM to scalar mesons is
straightforward, not involving any additional fit or alteration of the
interactions. The only approximation is the omission of color splitting, which
we verified to be negligible anyhow, due to the $^3P_0$ nature of the scalars
themselves. The thus obtained $S$-matrices for all mesonic $J^{PC}=0^{++}$
states have been searched for poles in the second Riemann sheet. The results
for the real parts of the found poles are given in Table~\ref{tab} \cite{B99b},
together with the predictions of model \cite{T95,TR96} (HUQM), the
experimentally confirmed candidates, and the respective interpretations in
terms of $q\bar{q}$ states. The most striking feature of the NUMM results is a
doubling of states
with respect to the bare $q\bar{q}$ spectrum, which was already forboded
somehow by the toy model presented above. Here, we restrict ourselves to note
that this dynamical phenomenon is typical for $S$-wave scattering channels
strongly coupled to confined channels, and refer to Refs.~\cite{B99b,B84} for
more details and discussion. The resonance doubling allows for an
identification of all observed scalar states up to 1.5 GeV, even obtaining one
extra resonance not yet confirmed by experiment, namely a light $K_0^*$ (old
$\kappa$). But also this state has recently received renewed phenomenological
and theoretical support \cite{I97,B98,R99,O99,N98}. As to the model of
Ref.~\cite{T95,TR96}, we observe from Table~\ref{tab} that such a resonance is
not predicted, nor the established $f_0$(1500). Moreover, the $f_0$(1300) is
interpreted as mainly $s\bar{s}$, while we claim it is predominantly $n\bar{n}$
and the $f_0$(1500) mainly $s\bar{s}$. Apart from considering our
interpretation more natural and favored by the known decay rates
\cite{B99b}, we should mention a very recent lattice calculation largely
supporting our model prediction \cite{W99}. The fact that model \cite{T95,TR96}
fails to find two complete scalar nonets we ascribe to the use of coupling
constants for the three-meson vertices that are not flavor independent
\cite{B99b}. Here, the crucial point is a point-particle approach in the
derivation of the couplings, which leads to a wrong normalization in the case
of the scalar mesons, being $^3P_0$ states themselves just as the created
$q\bar{q}$ pairs \cite{B99c,B99a}.

\begin{figure}[ht]
\normalsize
\begin{center}
\begin{picture}(283.46,180.08)(-50.00,-30.00)
%
%
%
%
\put(0.00,-5.52){\makebox(0,0)[tc]{0.6}}
%
\put(36.62,-5.52){\makebox(0,0)[tc]{0.8}}
%
\put(73.23,-5.52){\makebox(0,0)[tc]{1.0}}
%
\put(109.85,-5.52){\makebox(0,0)[tc]{1.2}}
%
\put(146.46,-5.52){\makebox(0,0)[tc]{1.4}}
%
\put(183.08,-5.52){\makebox(0,0)[tc]{1.6}}
%
%
%
%
%
%
\put(-5.52,32.70){\makebox(0,0)[rc]{100}}
%
\put(-5.52,65.40){\makebox(0,0)[rc]{200}}
%
\put(-5.52,98.10){\makebox(0,0)[rc]{300}}
%
%
%
%
\put(234.34,-5.52){\makebox(0,0)[tr]{GeV}}
\put(113.16,-20.07){\makebox(0,0)[tc]{$K\pi$ invariant mass}}
\put(-5.52,136.59){\makebox(0,0)[tr]{degrees}}
\put(5.52,132.57){\makebox(0,0)[tl]{phase}}
\put(23.80,6.87){\makebox(0,0){$\odot$}}
\put(32.95,8.89){\makebox(0,0){$\odot$}}
\put(40.28,13.31){\makebox(0,0){$\odot$}}
\put(44.85,12.03){\makebox(0,0){$\odot$}}
\put(46.68,13.21){\makebox(0,0){$\odot$}}
\put(48.52,12.43){\makebox(0,0){$\odot$}}
\put(50.35,12.46){\makebox(0,0){$\odot$}}
\put(52.18,12.79){\makebox(0,0){$\odot$}}
\put(54.01,12.95){\makebox(0,0){$\odot$}}
\put(55.84,11.35){\makebox(0,0){$\odot$}}
\put(57.67,10.95){\makebox(0,0){$\odot$}}
\put(59.50,12.16){\makebox(0,0){$\odot$}}
\put(61.33,13.37){\makebox(0,0){$\odot$}}
\put(63.16,14.00){\makebox(0,0){$\odot$}}
\put(64.99,15.34){\makebox(0,0){$\odot$}}
\put(69.57,15.83){\makebox(0,0){$\odot$}}
\put(76.89,17.62){\makebox(0,0){$\odot$}}
\put(84.22,18.90){\makebox(0,0){$\odot$}}
\put(91.54,19.68){\makebox(0,0){$\odot$}}
\put(98.86,20.01){\makebox(0,0){$\odot$}}
\put(106.19,21.68){\makebox(0,0){$\odot$}}
\put(113.51,22.66){\makebox(0,0){$\odot$}}
\put(120.83,23.44){\makebox(0,0){$\odot$}}
\put(128.15,25.93){\makebox(0,0){$\odot$}}
\put(135.11,29.43){\makebox(0,0){$\odot$}}
\put(141.70,31.95){\makebox(0,0){$\odot$}}
\put(149.57,36.62){\makebox(0,0){$\odot$}}
\put(157.26,40.22){\makebox(0,0){$\odot$}}
\put(164.77,45.61){\makebox(0,0){$\odot$}}
\put(175.76,56.90){\makebox(0,0){$\odot$}}
\put(192.23,74.88){\makebox(0,0){$\odot$}}
\put(210.54,122.29){\makebox(0,0){$\odot$}}
\put(228.85,135.37){\makebox(0,0){$\odot$}}
\put(41.92,9.16){\makebox(0,0){$\bullet$}}
\put(43.39,10.79){\makebox(0,0){$\bullet$}}
\put(45.77,12.10){\makebox(0,0){$\bullet$}}
\put(46.68,11.44){\makebox(0,0){$\bullet$}}
\put(48.52,13.41){\makebox(0,0){$\bullet$}}
\put(50.90,12.75){\makebox(0,0){$\bullet$}}
\put(54.01,13.41){\makebox(0,0){$\bullet$}}
\put(56.39,11.77){\makebox(0,0){$\bullet$}}
\put(58.04,14.06){\makebox(0,0){$\bullet$}}
\put(60.23,13.41){\makebox(0,0){$\bullet$}}
\put(61.33,14.39){\makebox(0,0){$\bullet$}}
\put(63.16,14.71){\makebox(0,0){$\bullet$}}
\put(67.37,15.04){\makebox(0,0){$\bullet$}}
\put(70.49,15.37){\makebox(0,0){$\bullet$}}
\put(75.06,17.00){\makebox(0,0){$\bullet$}}
\put(79.46,17.66){\makebox(0,0){$\bullet$}}
\put(86.96,18.31){\makebox(0,0){$\bullet$}}
\put(94.10,18.97){\makebox(0,0){$\bullet$}}
\put(101.06,19.62){\makebox(0,0){$\bullet$}}
\put(108.02,20.93){\makebox(0,0){$\bullet$}}
\put(115.34,22.89){\makebox(0,0){$\bullet$}}
\put(122.66,24.52){\makebox(0,0){$\bullet$}}
\put(129.99,27.47){\makebox(0,0){$\bullet$}}
\put(133.65,27.79){\makebox(0,0){$\bullet$}}
\put(137.31,30.74){\makebox(0,0){$\bullet$}}
\put(141.52,31.72){\makebox(0,0){$\bullet$}}
\put(145.00,34.66){\makebox(0,0){$\bullet$}}
\put(148.29,37.93){\makebox(0,0){$\bullet$}}
\put(151.95,42.18){\makebox(0,0){$\bullet$}}
\put(155.62,43.82){\makebox(0,0){$\bullet$}}
\put(160.01,45.78){\makebox(0,0){$\bullet$}}
\put(162.94,49.05){\makebox(0,0){$\bullet$}}
\put(167.15,50.36){\makebox(0,0){$\bullet$}}
\put(171.18,50.03){\makebox(0,0){$\bullet$}}
\put(175.02,52.64){\makebox(0,0){$\bullet$}}
\put(178.87,54.93){\makebox(0,0){$\bullet$}}
%
\end{picture}
\end{center}
\normalsize
\caption[]{Kaon-pion $I=\frac{1}{2}$ $S$-wave phase shifts. The data indicated
by $\odot$ are taken from Ref.~\cite{Est78} and by $\bullet$ from
Ref.~\cite{Aston}. The model results (dashed line) are taken from
Ref.~\cite{B86}. \\ 
Figure reprinted from: E.\ van Beveren and G.\ Rupp, \em Comment on
``Understanding the scalar meson $q\bar{q}$ nonet'', \em Eur.\ Phys.\ J. C
{\bf10}, Page no.\ 470, Copyright (1999), with permission from
Springer-Verlag.}
\label{kpis}
\end{figure}
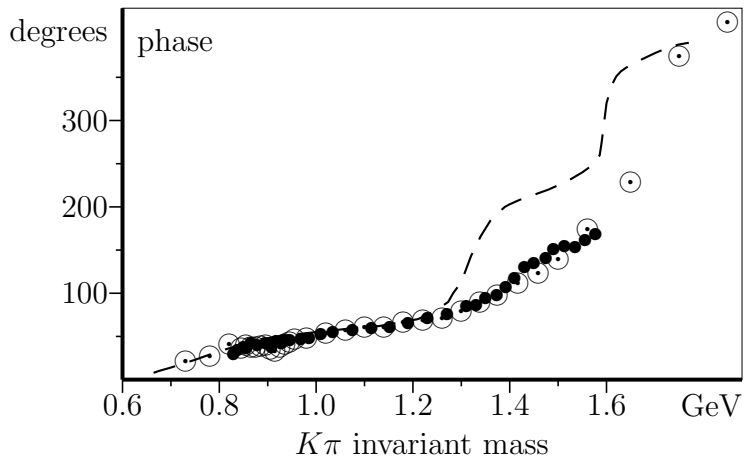

Another feature of the NUMM, and also of the HUQM for that matter, is the
automatic obtainment of a unitary, analytic $S$-matrix, which allows for a
straightforward calculation of partial cross sections and phase shifts. Thus,
we present our results for the elastic $S$-wave $\pi\pi$ and $K\pi$ phase
shifts, in Figs.~\ref{pipis} and \ref{kpis}, respectively. We must reemphasize
that these are model \em predictions \em \/ and not the result of a fit. In
that perspective, the results are surprisingly good, reproducing the bulk
features of the experimental phases, including the resonant structures. In the
$\pi\pi$ case, both the broad structure from roughly $400$ to $950$ MeV, owing
to a $f_0$(400-1200) (or $\sigma$) pole at $470-208i$ MeV, and the sharp
resonance close to 1 GeV, due to a $f_0$(980) pole at $994-17i$ MeV, are
reasonably well described. Of course, some background structure is clearly
lacking, which is no surprise in view of the neglect of final-state
interactions in the meson-meson channels. Also notice that in the $K\pi$ case,
where final-state interactions from $t$-channel meson exchanges are expected to
be less important, the phase shifts are extremely well reproduced in the energy
region 0.7--1.2 GeV, exactly where we find the lowest $K_0^*$ pole, i.e., at
$727-263i$ MeV. So it is unmistakably demonstrated that this very controversial
resonance is perfectly compatible with a non-resonant behavior of the phases in
the same region.
\section*{Concluding remarks}
In the foregoing, we hope to have made it clear that, for a reliable and
detailed description of meson spectra in general, the effects of mesonic loops
and decay should be taken into account. In partucular, this is \em a forteriori
\em \/ true in the case of the scalar mesons, where, from a theoretical point
of view, one is faced with very large couplings to $S$-wave two-meson channels,
and, on the experimental side, a confusing picture of many broad as well as
narrow resonances of very disparate masses emerges. Whatever the used approach,
however, extreme care is required in the choice of the classes of the to be
included decay channels, and in the computation of the respective coupling
constants, lest one introduce explicitly flavor-breaking mechanisms that
may distort the spectra in an unrealistic fashion.

The NUMM employed here may, of course, be subject to improvements. As indicated
before, final-state interactions from $t$-channel meson exchanges should be
included, aiming at restoring crossing symmetry to some degree, and in order to
allow for a more accurate reproduction of the experimental meson-meson phase
shifts. Furthermore, relativity should be addressed in a thorougher way than
just by some relativistic kinematics, preferably in a covariant quasipotential
framework. This would require a profound overhaul of the mathematical
formulation of the model, but could then make further refinements of the used
interactions feasible.
\begin{table}[t]
\caption[]{Scalar-meson predictions and $q\bar{q}$ interpretations for the HUQM
and NUMM, together with experimentally established states. \\
Table reprinted from: E.\ van Beveren and G.\ Rupp, \em Comment on
``Understanding the scalar meson $q\bar{q}$ nonet'', \em Eur.\ Phys.\ J. C
{\bf10}, Page no.\ 471, Copyright (1999), with permission from
Springer-Verlag.}
\begin{tabular}{c||c|l||c|l||c}
& \multicolumn{2}{c||}{HUQM \cite{T95,TR96}} &
\multicolumn{2}{c||}{NUMM \cite{B86,B99b}} &
Exp.\ \cite{PDG98} \\ \hline 
Resonance & \mbox{Re}$E_{\mbox{\scriptsize pole}}$ & $q\bar{q}$ configuration &
\mbox{Re}$E_{\mbox{\scriptsize pole}}$ & $q\bar{q}$ configuration & Mass \\
\hline
$\sigma /f_0$(400--1200) & 470 & $1^{\textstyle st}$
$\approx \frac{1}{\sqrt{2}}(u\bar{u}+d\bar{d})$
& 470  & $1^{\textstyle st}$
$\approx \frac{1}{\sqrt{2}}(u\bar{u}+d\bar{d})$ & 400-1200 \\
$S^{\ast}/f_0$(980) & 1006 & $1^{\textstyle st}$ $\approx s\bar{s}$
& 994  & $1^{\textstyle st}$ $\approx s\bar{s}$ & 980 $\pm$ 10 \\
$\delta /a_0$(980) & 1094 & $1^{\textstyle st}$
$\frac{1}{\sqrt{2}}(u\bar{u}-d\bar{d})$
& 968  & $1^{\textstyle st}$
$\frac{1}{\sqrt{2}}(u\bar{u}-d\bar{d})$ & 983 $\pm$ 1 \\
$\kappa /K_0^*$ & - & \hspace{1.5cm} -
& 727  &  $1^{\textstyle st}$ $s\bar{d}$ & ? \\
$f_0$(1370) & 1214 & $2^{\textstyle nd}$ $\approx s\bar{s}$             
& 1300 & $2^{\textstyle nd}$
$ \approx\frac{1}{\sqrt{2}}(u\bar{u}+d\bar{d})$& 1200--1500 \\
$f_0$(1500) &  - & \hspace{1.5cm} - 
& 1500 & $2^{\textstyle nd}$ $\approx s\bar{s}$ & 1500 $\pm$ 10 \\
$a_0$(1450) & 1592 & $2^{\textstyle nd}$
$\frac{1}{\sqrt{2}}(u\bar{u}-d\bar{d})$  
&1300  & $2^{\textstyle nd}$
$\frac{1}{\sqrt{2}}(u\bar{u}-d\bar{d})$ & 1474 $\pm$ 19 \\
$K_0^*$(1430) & 1450 & $1^{\textstyle st}$
$s\bar{d}$ & 1400 & $2^{\textstyle nd}$ $s\bar{d}$ 
& 1429 $\pm$ 6
\end{tabular}
\label{tab}
\end{table}
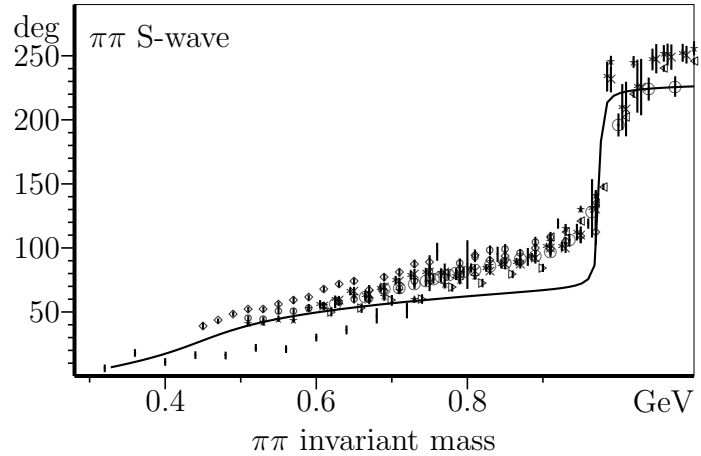
\begin{figure}[ht]
\normalsize
\begin{center}
\begin{picture}(283.46,180.08)(-50.00,-30.00)
%
%
%
%
\put(34.29,-5.52){\makebox(0,0)[tc]{0.4}}
%
\put(91.45,-5.52){\makebox(0,0)[tc]{0.6}}
%
\put(148.61,-5.52){\makebox(0,0)[tc]{0.8}}
%
%
%
%
%
\put(-5.52,24.24){\makebox(0,0)[rc]{50}}
%
\put(-5.52,48.48){\makebox(0,0)[rc]{100}}
%
\put(-5.52,72.73){\makebox(0,0)[rc]{150}}
%
\put(-5.52,96.97){\makebox(0,0)[rc]{200}}
%
\put(-5.52,121.21){\makebox(0,0)[rc]{250}}
%
%
%
%
%
%
%
%
%
%
%
%
%
%
%
%
%
%
%
%
\put(234.34,-5.52){\makebox(0,0)[tr]{GeV}}
\put(113.16,-20.07){\makebox(0,0)[tc]{$\pi\pi$ invariant mass}}
\put(-5.52,136.59){\makebox(0,0)[tr]{deg}}
\put(5.52,132.57){\makebox(0,0)[tl]{$\pi \pi$ S-wave}}
\tiny
\put(11.43,2.91){\makebox(0,0){$\cdot$}}
%
\put(22.86,8.73){\makebox(0,0){$\cdot$}}
%
\put(34.29,5.33){\makebox(0,0){$\cdot$}}
%
\put(45.72,7.90){\makebox(0,0){$\cdot$}}
%
\put(57.16,7.76){\makebox(0,0){$\cdot$}}
%
\put(68.59,10.67){\makebox(0,0){$\cdot$}}
%
\put(80.02,10.18){\makebox(0,0){$\cdot$}}
%
\put(91.45,14.55){\makebox(0,0){$\cdot$}}
%
\put(102.88,17.45){\makebox(0,0){$\cdot$}}
%
\put(114.31,22.79){\makebox(0,0){$\cdot$}}
%
\put(125.74,24.73){\makebox(0,0){$\cdot$}}
%
\put(137.17,47.03){\makebox(0,0){$\cdot$}}
%
\put(148.61,42.18){\makebox(0,0){$\cdot$}}
%
\put(160.04,46.06){\makebox(0,0){$\cdot$}}
%
\put(171.47,45.09){\makebox(0,0){$\cdot$}}
%
\put(182.90,57.70){\makebox(0,0){$\cdot$}}
%
\put(194.33,57.70){\makebox(0,0){$\cdot$}}
%
\put(65.73,21.77){\makebox(0,0){$\circ$}}
%
\put(71.45,21.38){\makebox(0,0){$\circ$}}
%
\put(77.16,24.58){\makebox(0,0){$\circ$}}
%
\put(82.88,24.44){\makebox(0,0){$\circ$}}
%
\put(88.59,25.65){\makebox(0,0){$\circ$}}
%
\put(94.31,26.47){\makebox(0,0){$\circ$}}
%
\put(100.02,27.98){\makebox(0,0){$\circ$}}
%
\put(105.74,29.24){\makebox(0,0){$\circ$}}
%
\put(111.45,29.14){\makebox(0,0){$\circ$}}
%
\put(117.17,32.14){\makebox(0,0){$\circ$}}
%
\put(122.89,33.31){\makebox(0,0){$\circ$}}
%
\put(128.60,35.93){\makebox(0,0){$\circ$}}
%
\put(134.32,39.47){\makebox(0,0){$\circ$}}
%
\put(145.75,42.91){\makebox(0,0){$\circ$}}
%
\put(151.46,44.90){\makebox(0,0){$\circ$}}
%
\put(157.18,47.56){\makebox(0,0){$\circ$}}
%
\put(162.89,48.14){\makebox(0,0){$\circ$}}
%
\put(168.61,46.64){\makebox(0,0){$\circ$}}
%
\put(174.33,50.71){\makebox(0,0){$\circ$}}
%
\put(180.04,52.51){\makebox(0,0){$\circ$}}
%
\put(92.88,27.15){\makebox(0,0){$\ast$}}
%
\put(98.59,29.05){\makebox(0,0){$\ast$}}
%
\put(104.31,32.04){\makebox(0,0){$\ast$}}
%
\put(110.03,30.62){\makebox(0,0){$\ast$}}
%
\put(115.74,33.15){\makebox(0,0){$\ast$}}
%
\put(121.46,36.30){\makebox(0,0){$\ast$}}
%
\put(127.17,38.20){\makebox(0,0){$\ast$}}
%
\put(132.89,39.31){\makebox(0,0){$\ast$}}
%
\put(138.60,38.99){\makebox(0,0){$\ast$}}
%
\put(144.32,37.89){\makebox(0,0){$\ast$}}
%
\put(150.03,40.89){\makebox(0,0){$\ast$}}
%
\put(155.75,40.89){\makebox(0,0){$\ast$}}
%
\put(161.47,42.31){\makebox(0,0){$\ast$}}
%
\put(167.18,43.25){\makebox(0,0){$\ast$}}
%
\put(172.90,45.15){\makebox(0,0){$\ast$}}
%
\put(178.61,49.25){\makebox(0,0){$\ast$}}
%
\put(184.33,49.10){\makebox(0,0){$\ast$}}
%
\put(190.04,54.46){\makebox(0,0){$\ast$}}
%
\put(195.76,63.46){\makebox(0,0){$\ast$}}
%
\put(201.48,113.34){\makebox(0,0){$\ast$}}
%
\put(207.19,101.82){\makebox(0,0){$\ast$}}
%
\put(212.91,109.71){\makebox(0,0){$\ast$}}
%
\put(218.62,119.81){\makebox(0,0){$\ast$}}
%
\put(224.34,121.70){\makebox(0,0){$\ast$}}
%
\put(230.05,122.34){\makebox(0,0){$\ast$}}
%
\put(65.73,20.02){\makebox(0,0){$\star$}}
%
\put(71.45,20.22){\makebox(0,0){$\star$}}
%
\put(77.16,21.67){\makebox(0,0){$\star$}}
%
\put(82.88,21.33){\makebox(0,0){$\star$}}
%
\put(88.59,25.79){\makebox(0,0){$\star$}}
%
\put(94.31,26.86){\makebox(0,0){$\star$}}
%
\put(100.02,28.31){\makebox(0,0){$\star$}}
%
\put(105.74,30.45){\makebox(0,0){$\star$}}
%
\put(111.45,29.62){\makebox(0,0){$\star$}}
%
\put(117.17,30.06){\makebox(0,0){$\star$}}
%
\put(122.89,32.53){\makebox(0,0){$\star$}}
%
\put(128.60,28.85){\makebox(0,0){$\star$}}
%
\put(134.32,35.73){\makebox(0,0){$\star$}}
%
\put(140.03,34.76){\makebox(0,0){$\star$}}
%
\put(145.75,36.07){\makebox(0,0){$\star$}}
%
\put(151.46,37.82){\makebox(0,0){$\star$}}
%
\put(157.18,41.70){\makebox(0,0){$\star$}}
%
\put(162.89,43.64){\makebox(0,0){$\star$}}
%
\put(180.04,50.96){\makebox(0,0){$\star$}}
%
\put(185.76,55.95){\makebox(0,0){$\star$}}
%
\put(191.47,63.13){\makebox(0,0){$\star$}}
%
\put(197.19,68.46){\makebox(0,0){$\star$}}
%
\put(202.90,119.32){\makebox(0,0){$\star$}}
%
\put(211.48,118.88){\makebox(0,0){$\star$}}
%
\put(222.91,122.13){\makebox(0,0){$\star$}}
%
\put(234.34,124.02){\makebox(0,0){$\star$}}
%
\put(94.31,26.33){\makebox(0,0){$\times$}}
%
\put(100.02,28.58){\makebox(0,0){$\times$}}
%
\put(105.74,32.19){\makebox(0,0){$\times$}}
%
\put(111.45,31.08){\makebox(0,0){$\times$}}
%
\put(117.17,33.62){\makebox(0,0){$\times$}}
%
\put(122.89,36.02){\makebox(0,0){$\times$}}
%
\put(128.60,38.47){\makebox(0,0){$\times$}}
%
\put(134.32,39.37){\makebox(0,0){$\times$}}
%
\put(140.03,39.34){\makebox(0,0){$\times$}}
%
\put(145.75,38.18){\makebox(0,0){$\times$}}
%
\put(151.46,39.71){\makebox(0,0){$\times$}}
%
\put(157.18,39.88){\makebox(0,0){$\times$}}
%
\put(162.89,42.25){\makebox(0,0){$\times$}}
%
\put(168.61,42.50){\makebox(0,0){$\times$}}
%
\put(174.33,44.73){\makebox(0,0){$\times$}}
%
\put(180.04,48.41){\makebox(0,0){$\times$}}
%
\put(185.76,48.65){\makebox(0,0){$\times$}}
%
\put(191.47,53.45){\makebox(0,0){$\times$}}
%
\put(197.19,63.01){\makebox(0,0){$\times$}}
%
\put(202.90,112.34){\makebox(0,0){$\times$}}
%
\put(208.62,101.07){\makebox(0,0){$\times$}}
%
\put(214.34,109.38){\makebox(0,0){$\times$}}
%
\put(220.05,120.10){\makebox(0,0){$\times$}}
%
\put(225.77,120.73){\makebox(0,0){$\times$}}
%
\put(231.48,121.62){\makebox(0,0){$\times$}}
%
\put(48.58,18.96){\makebox(0,0){$\diamond$}}
%
\put(54.30,20.90){\makebox(0,0){$\diamond$}}
%
\put(60.01,23.51){\makebox(0,0){$\diamond$}}
%
\put(65.73,25.11){\makebox(0,0){$\diamond$}}
%
\put(71.45,25.21){\makebox(0,0){$\diamond$}}
%
\put(77.16,27.10){\makebox(0,0){$\diamond$}}
%
\put(82.88,28.65){\makebox(0,0){$\diamond$}}
%
\put(88.59,29.72){\makebox(0,0){$\diamond$}}
%
\put(94.31,32.87){\makebox(0,0){$\diamond$}}
%
\put(100.02,34.71){\makebox(0,0){$\diamond$}}
%
\put(105.74,35.88){\makebox(0,0){$\diamond$}}
%
\put(111.45,32.63){\makebox(0,0){$\diamond$}}
%
\put(117.17,37.28){\makebox(0,0){$\diamond$}}
%
\put(122.89,39.08){\makebox(0,0){$\diamond$}}
%
\put(128.60,42.42){\makebox(0,0){$\diamond$}}
%
\put(134.32,43.15){\makebox(0,0){$\diamond$}}
%
\put(145.75,40.63){\makebox(0,0){$\diamond$}}
%
\put(151.46,45.82){\makebox(0,0){$\diamond$}}
%
\put(157.18,43.78){\makebox(0,0){$\diamond$}}
%
\put(162.89,45.33){\makebox(0,0){$\diamond$}}
%
\put(168.61,43.25){\makebox(0,0){$\diamond$}}
%
\put(174.33,48.39){\makebox(0,0){$\diamond$}}
%
\put(180.04,48.05){\makebox(0,0){$\diamond$}}
%
\put(185.76,51.20){\makebox(0,0){$\diamond$}}
%
\put(191.47,52.70){\makebox(0,0){$\diamond$}}
%
\put(197.19,54.50){\makebox(0,0){$\diamond$}}
%
\put(180.04,52.36){\makebox(0,0){$\triangleleft$}}
%
\put(185.76,54.30){\makebox(0,0){$\triangleleft$}}
%
\put(191.47,58.67){\makebox(0,0){$\triangleleft$}}
%
\put(197.19,65.45){\makebox(0,0){$\triangleleft$}}
%
\put(200.05,71.66){\makebox(0,0){$\triangleleft$}}
%
\put(208.62,97.70){\makebox(0,0){$\triangleleft$}}
%
\put(211.48,106.57){\makebox(0,0){$\triangleleft$}}
%
\put(222.91,116.36){\makebox(0,0){$\triangleleft$}}
%
\put(234.34,119.13){\makebox(0,0){$\triangleleft$}}
%
\put(97.17,24.29){\makebox(0,0){$\triangleright$}}
%
\put(108.60,25.70){\makebox(0,0){$\triangleright$}}
%
\put(120.03,28.61){\makebox(0,0){$\triangleright$}}
%
\put(131.46,29.14){\makebox(0,0){$\triangleright$}}
%
\put(142.89,33.41){\makebox(0,0){$\triangleright$}}
%
\put(154.32,35.20){\makebox(0,0){$\triangleright$}}
%
\put(165.75,38.50){\makebox(0,0){$\triangleright$}}
%
\put(177.18,40.92){\makebox(0,0){$\triangleright$}}
%
\put(98.59,27.15){\makebox(0,0){$\odot$}}
%
\put(110.03,30.06){\makebox(0,0){$\odot$}}
%
\put(117.17,32.97){\makebox(0,0){$\odot$}}
%
\put(122.89,33.45){\makebox(0,0){$\odot$}}
%
\put(128.60,34.91){\makebox(0,0){$\odot$}}
%
\put(132.89,35.88){\makebox(0,0){$\odot$}}
%
\put(135.75,36.85){\makebox(0,0){$\odot$}}
%
\put(138.60,37.33){\makebox(0,0){$\odot$}}
%
\put(141.46,38.30){\makebox(0,0){$\odot$}}
%
\put(144.32,38.79){\makebox(0,0){$\odot$}}
%
\put(147.18,39.27){\makebox(0,0){$\odot$}}
%
\put(151.46,40.24){\makebox(0,0){$\odot$}}
%
\put(157.18,41.70){\makebox(0,0){$\odot$}}
%
\put(162.89,42.67){\makebox(0,0){$\odot$}}
%
\put(168.61,44.12){\makebox(0,0){$\odot$}}
%
\put(174.33,45.58){\makebox(0,0){$\odot$}}
%
\put(180.04,47.03){\makebox(0,0){$\odot$}}
%
\put(187.19,51.39){\makebox(0,0){$\odot$}}
%
\put(195.76,62.06){\makebox(0,0){$\odot$}}
%
\put(205.76,95.03){\makebox(0,0){$\odot$}}
%
\put(217.19,108.60){\makebox(0,0){$\odot$}}
%
\put(227.20,109.57){\makebox(0,0){$\odot$}}
%
\normalsize
%
\end{picture}
\end{center}
\normalsize
\caption[]{Model \cite{B86} results for $\pi\pi$ elastic $S$-wave phase shifts.
The various sets of data are taken from ($\odot$, \cite{Pro73}), ($\ast$,
\cite{Hya73}), ($\star$, $\times$, $\diamond$, $\triangleleft$,
$\triangleright$ respectively for analyses A, B, C, D, and E of
\cite{Gra74}), ($\circ$, \cite{Est74}), and ($\cdot$, \cite{Bis81}). \\
Figure reprinted from: E.\ van Beveren, \em Scalar mesons as $q\bar{q}$ systems
with meson-meson admixtures, \em Nucl.\ Phys.\ B (Proc.\ Suppl.) {\bf21}, Page
no.\ 47, Copyright (1991), with permission from Elsevier Science.}
\label{pipis}
\end{figure}

\end{document}